\documentclass[12pt]{article}
\usepackage{axodraw,bbold}

\catcode`@=12
\begin{document}
\thispagestyle{empty}
\begin{flushright} 
UCRHEP-T383\\ 
April 2005\
\end{flushright}
\vspace{0.5in}
\begin{center}
{\LARGE	\bf Connection between the neutrino seesaw\\ mechanism and
properties of the\\ Majorana neutrino mass matrix\\}
\vspace{1.5in}
{\bf Ernest Ma\\}
\vspace{0.2in}
{\sl Physics Department, University of California, Riverside, 
California 92521, USA\\}
\vspace{1.5in}
\end{center}

\begin{abstract}\
If it can be ascertained experimentally that the $3 \times 3$ Majorana 
neutrino mass matrix ${\cal M}_\nu$ has vanishing determinants for one 
or more of its $2 \times 2$ submatrices, it may be interpreted as supporting 
evidence for the theoretically well-known canonical seesaw mechanism. 
I show how these two things are connected and offer a realistic ${\cal M}_\nu$ 
with two zero subdeterminants as an example.
\end{abstract}

\newpage
\baselineskip 24pt

It is a common theoretical belief in neutrino physics that the observed 
smallness of neutrino masses is due to the celebrated canonical seesaw 
mechanism \cite{seesaw}, i.e.
\begin{equation}
{\cal M}_\nu = {\cal M}_D {\cal M}_N^{-1} {\cal M}_D^T.
\end{equation}
Here ${\cal M}_D$ is the $3 \times 3$ Dirac mass matrix linking $(\nu_e, 
\nu_\mu, \nu_\tau)$ to their right-handed singlet counterparts $(N_e, N_\mu, 
N_\tau)$, and ${\cal M}_N$ is the $3 \times 3$ Majorana mass matrix of the 
latter.  Assuming ${\cal M}_D$ to be of order the electroweak breaking scale, 
a very large ${\cal M}_N$ would then result in a very small ${\cal M}_\nu$. 
However, it is impossible to verify this hypothesis without reaching 
energies of the scale of ${\cal M}_N$, or extreme 
sensitivities in rare decay processes.  Both are hopeless 
in the near future unless ${\cal M}_N$ is of order a few TeV \cite{ma01}. 
If ${\cal M}_N$ is much greater than that, one may never know if Eq.~(1) 
is really how neutrinos become massive.

The form and texture of ${\cal M}_\nu$ have been under theoretical study 
for many years.  Is it possible at all to discover from its structure 
that it actually comes from ${\cal M}_N$ as given by Eq.~(1)?  The answer 
is ``yes'', provided that ${\cal M}_N$ has one or more texture zeros. 
In that case, ${\cal M}_N^{-1}$ has one or more $2 \times 2$ submatrices 
with zero determinants.  If ${\cal M}_D$ is also diagonal, this property is 
preserved in ${\cal M}_\nu$.  Finding such a structure in the latter 
experimentally would be provocative supporting evidence that Eq.~(1) is 
correct!

In the basis where the charged-lepton mass matrix ${\cal M}_l$ is diagonal, 
the possible existence of texture zeros in ${\cal M}_\nu$ have been 
considered previously \cite{fgm02}.  These zeros are derivable from 
Abelian discrete symmetries \cite{gjlt04}, and in the case of 
$({\cal M}_\nu)_{\mu \mu} = ({\cal M}_\nu)_{\tau \tau} = 0$ also from 
the non-Abelian discrete groups $Q_8$ \cite{fkmt04} and $D_5$ \cite{ma04}. 
However, if ${\cal M}_N$ is the progenitor of ${\cal M}_\nu$, one 
should perhaps consider instead the texture zeros of the former \cite{l04}, 
which may be similarly obtained from the symmetries already mentioned. 
For example, if $({\cal M}_N)_{\mu \mu} = ({\cal M}_N)_{\tau \tau} = 0$, i.e.
\begin{equation}
{\cal M}_N = \pmatrix{A & B & C \cr B & 0 & D \cr C & D & 0},
\end{equation}
which is the analog of Scenario (1) of Ref.~\cite{fkmt04} and also that of 
the model of Ref.~\cite{ma04}, then
\begin{equation}
{\cal M}_N^{-1} = \pmatrix{a & b & c \cr b & e & d \cr c & d & f}
\end{equation}
has two zero $2 \times 2$ determinants, i.e.
\begin{equation}
ae - b^2 = af - c^2 = 0.
\end{equation}
[To prove this, one simply considers the identity ${\cal M}_N^{-1} {\cal M}_N 
= 1$.]

If ${\cal M}_D$ is also diagonal, which may be maintained again by the 
symmetries already mentioned \cite{fkmt04,ma04}, i.e.
\begin{equation}
{\cal M}_D = \pmatrix{x & 0 & 0 \cr 0 & y & 0 \cr 0 & 0 & z},
\end{equation}
then from Eq.~(1),
\begin{equation}
{\cal M}_\nu = \pmatrix{x^2 a & xyb & xzc \cr xyb & y^2 e & yzd \cr xzc & 
yzd & z^2 f}.
\end{equation}
Since
\begin{eqnarray}
(x^2 a)(y^2 e) - (xyb)^2 &=& x^2 y^2 (ae - b^2), \\
(x^2 a)(z^2 f) - (xzc)^2 &=& x^2 z^2 (af - c^2),
\end{eqnarray}
the corresponding two subdeterminants of ${\cal M}_\nu$ are zero as well. 
Using Eqs.~(4), (7) and (8), let us rewrite ${\cal M}_\nu$ of Eq.~(6) as
\begin{equation}
{\cal M}_\nu = \pmatrix{\alpha & \beta & \gamma \cr \beta & 
\alpha^{-1} \beta^2 & \delta \cr \gamma & \delta & \alpha^{-1} \gamma^2}.
\end{equation}
This is Model (D) of Ref.~\cite{l04}. It is in fact a realistic neutrino 
mass matrix, capable of describing all present data \cite{data}.  If 
confirmed by future precision data, this would be provocative supporting 
evidence that the long-held theoretical belief in the canonical seesaw 
mechanism is indeed valid!

The ${\cal M}_\nu$ of Eq.~(9) has four parameters, but it will fit all data 
even if it is reduced to three parameters by setting $\beta=\gamma$, i.e.
\begin{equation}
{\cal M}_\nu = \pmatrix{\alpha & \beta & \beta \cr \beta & 
\alpha^{-1}\beta^2 & \delta \cr \beta & \delta & \alpha^{-1}\beta^2}.
\end{equation}
This is a special case of the general form \cite{ma02} which exhibits the 
symmetry $\nu_\mu \leftrightarrow \nu_\tau$, implying $\theta_{23} = \pi/4$ 
and $\theta_{13} = 0$ in the mixing matrix linking $\nu_e,\nu_\mu,\nu_\tau$ 
to their mass eigenstates.  Using the general analysis of Ref.~\cite{ma02}, 
where ${\cal M}_\nu$ is given by
\begin{equation}
{\cal M}_\nu = \pmatrix{a+2b+2c & d & d \cr d & b & a+b \cr d & a+b & b},
\end{equation}
we then have
\begin{equation}
d = \beta, ~~~ b = \alpha^{-1}\beta^2, ~~~ a = \delta - \alpha^{-1}\beta^2, 
~~~ c = (\alpha - \delta - \alpha^{-1}\beta^2)/2.
\end{equation}
As shown in Ref.~\cite{ma02}, ${\cal M}_\nu$ is exactly diagonalized by
\begin{equation}
\pmatrix{\nu_e \cr \nu_\mu \cr \nu_\tau} = \pmatrix{\cos \theta & 
-\sin \theta & 0 \cr \sin \theta/\sqrt 2 & \cos \theta/\sqrt 2 & -1/\sqrt 2 
\cr \sin \theta/\sqrt 2 & \cos \theta/\sqrt 2 & 1/\sqrt 2} \pmatrix{\nu_1 \cr 
\nu_2 \cr \nu_3},
\end{equation}
with
\begin{equation}
m_{1,2} = a+2b+c \mp \sqrt{c^2 + 2d^2}, ~~~ m_3 = -a, ~~~ \tan^2 2 \theta = 
2d^2/c^2.
\end{equation}
Using Eqs.~(12) and (14), the three parameters $\alpha, \beta, \delta$ of 
Eq.~(10) can now be fixed by the three experimental measurements of 
$\theta (= \theta_{12})$,
\begin{equation}
\Delta m_{sol}^2 = 4(a+2b+c)\sqrt{c^2+2d^2} = {4(a+2b+c)|c| \over 
\cos 2 \theta_{12}},
\end{equation}
and
\begin{equation}
\Delta m_{atm}^2 = a^2 - (a+2b+c)^2 - c^2 - 2d^2 = a^2 - \left[ 
{(\Delta m_{sol}^2) \cos 2 \theta_{12} \over 4c} \right]^2 - {c^2 \over 
\cos^2 2 \theta_{12}}.
\end{equation}
For example, let $\alpha = 6 \times 10^{-4}$ eV, $\beta = 4 \times 10^{-3}$ 
eV, and $\delta = -2.1 \times 10^{-2}$ eV, then Eq.~(14) yields a normal 
ordering of neutrino masses $(|m_1| < |m_2| < |m_3|)$, with
\begin{equation}
\tan^2 \theta_{12} = 0.42, ~~~ \Delta m^2_{sol} = 7.8 \times 10^{-5}~
{\rm eV}^2, ~~~  \Delta m^2_{atm} = 2.2 \times 10^{-3}~{\rm eV}^2,
\end{equation}
in good agreement with present data.

If $\theta_{13} \neq 0$ is required by future data, the unrestricted 
Eq.~(9) itself should be considered.  Instead of 
$({\cal M}_N)_{\mu \mu} = ({\cal M}_N)_{\tau \tau} = 0$ in Eq.~(2), 
another interesting possibility is to have $({\cal M}_N)_{\mu \tau} = 
({\cal M}_N)_{\tau \mu} = 0$.  In that case, ${\cal M}_N^{-1}$ of 
Eq.~(3) has $ad - bc = 0$.  This has in fact been implemented in a model 
\cite{gl03} based on $D_4 \times Z_2$.

Experimentally, it will be a daunting task to measure each of the six 
elements of ${\cal M}_\nu$.  Only the absolute value of $({\cal M}_\nu)_{ee}$ 
is subject to direct experimental measurement from neutrinoless double beta 
decay, which is being pursued vigorously by several international 
collaborations.  The absolute values of $({\cal M}_\nu)_{e\mu}$ and 
$({\cal M}_\nu)_{\mu\mu}$ may be obtained from future experiments searching 
for $\mu^-$ to $e^+$ and $\mu^-$ to $\mu^+$ conversion in nuclei, but the 
sensitivity required is many orders of magnitude beyond present capability. 
However, a partial test of the idea of zero subdeterminants is possible 
because such a requirement reduces the number of independent parameters in 
${\cal M}_\nu$.  If both $|m_{\nu_e}|$ and $|({\cal M}_\nu)_{ee}|$ are 
measured in the future, as well as the $CP$-nonconserving Dirac phase of the 
neutrino mixing matrix and its three angles, together with more precise 
values of $\Delta m^2_{atm}$ and $\Delta m^2_{sol}$, these eight quantities 
can be used to check if Eq.~(9) [or any of the other possible forms of 
${\cal M}_\nu$ with one or more zero subdeterminants] is still valid. 
If so, then it is at least indirect confirmation of this hypothesis. 
[Note that two zero subdeterminants imply four real parameters and one 
phase, and one zero subdeterminant implies five real parameters and two 
phases.]

Naturally small Majorana neutrino masses are obtainable in the Standard 
Model in three and only three tree-level mechanisms \cite{ma98}.  The 
canonical seesaw mechanism using heavy Majorana right-handed neutrino 
singlets $N_i$ has dominated the literature, but the use of a heavy Higgs 
scalar triplet $\xi$ without $N_i$ is just as natural \cite{ms98}.  In the 
latter case, ${\cal M}_\nu$ is obtained directly through the naturally 
small vacuum expectation value of $\xi$, and it makes sense to consider 
the possible texture zeros of ${\cal M}_\nu$ which may be derived from 
some discrete family symmetry \cite{fkmt04,ma04}.  On the other hand, if 
${\cal M}_N$ is truly the progenitor of ${\cal M}_\nu$ as dictated by 
Eq.~(1), then it makes more sense to consider the structure of ${\cal M}_N$ 
for its imprint on ${\cal M}_\nu$.  This is indeed possible 
if ${\cal M}_N$ has one or more texture zeros \cite{l04} and ${\cal M}_D$ 
is diagonal.  It is pointed out in this paper that a simple way to know is 
to look for zero subdeterminants in ${\cal M}_\nu$.  Finding them would 
go a long way in bolstering the neutrino community's faith in the correctness 
of Eq.~(1).  Present data are in fact consistent with such a prediction, 
as exemplified by Eqs. (9) and (10).  [If the neutrino mass matrix comes 
from a different mechanism, a zero subdeterminant may also occur 
accidentally, so there can never be a decisive proof.] Just as the original 
observation which led to the canonical seesaw mechanism is essentially 
trivial, the present observation that a zero entry in a matrix is reflected 
by a zero determinant in its inverse is also essentially trivial, but both 
may in fact be important clues to the origin of the observed neutrino mass 
matrix.

This work was supported in part by the U.~S.~Department of Energy under 
Grant No. DE-FG03-94ER40837.

\bibliographystyle{unsrt}

\begin{thebibliography}{99}
\bibitem{l04} L. Lavoura, Phys. Lett. {\bf B609}, 317 (2005).  This paper 
considers the texture zeros of ${\cal M}_\nu^{-1}$ which are equivalent to 
those of ${\cal M}_N$ if ${\cal M}_D$ is also diagonal.
\end{thebibliography}

\end{document}